\documentclass[twocolumn,a4paper,10pt]{article}
\pdfoutput=1
\usepackage{amsmath,amsfonts}
\usepackage{amssymb}
\usepackage{epsfig}
\usepackage{times}
\usepackage{epsfig}
\usepackage[table]{xcolor}
\usepackage{array}
\usepackage{graphicx}
\usepackage{subfiles}
\usepackage{multirow}
\usepackage{caption}
\usepackage{subcaption}
\usepackage[backend=biber,
style=authoryear,
sorting=nyt, 
natbib]{biblatex}
\makeatletter  

\newcounter{numbersec}
\renewcommand{\section}[1]{\par\noindent\stepcounter{numbersec}
	\par
	\vspace{6pt}
	\noindent\textbf{\large   \arabic{numbersec} \hspace*{0.3cm} #1 }
	\par
	\vspace{2pt}
}
\renewcommand{\subsection}[1]{
	\par
	\vspace{6pt}
	\noindent\textbf{#1}
	\par
}
\renewcommand{\subsubsection}[1]{%
	\par
	\vspace{6pt}
	\textbf{#1.}
}

\newcommand{\Abstract}{\par\vspace{6pt}\noindent\textbf{\large Abstract}\par\vspace{2pt}}
\newcommand{\Acknowledgments}{\par\vspace{6pt}\noindent\textbf{\large Acknowledgments }\par\vspace{2pt}}

\newenvironment{References}{
\par\vspace{6pt}\noindent\textbf{\large References}\par\vspace{2pt}
\begin{small}\begin{list}{ }{\itemsep2mm \parsep0mm\labelsep0mm\leftmargin0mm}}
{\end{list}\end{small}}
\newtheorem{problem}{Problem Statement}

\makeatother

\title{\vspace*{-12mm}
\LARGE \sc \textbf{  
Towards Long-Term Predictions of Turbulence \\ Using Neural Operators
}}

\author{ \Large \bf \textit{ 
F.A. Gonzalez$^{1,2}$, FX. Demoulin$^{1,2}$ and S. Bernard$^{1,3}$}  \\ \\
\bf  $^{1}$ \textit{University of Rouen - UFR Sciences et Techniques, Saint-Etienne du Rouvray,, France} \\
\bf  $^{2}$ \textit{CNRS UMR6614-C.O.R.I.A., Saint-Etienne du Rouvray, France} \\
\bf  $^{3}$ \textit{LITIS Lab, , Saint-Etienne du Rouvray, France} \\ \\
{\it fernando.gonzalez@coria.fr}
}
\date{}

\begin{document}
\maketitle
\thispagestyle{empty}

%%%%%%%%%%%%%%%%%%%%
% Paper text
%%%%%%%%%%%%%%%%%%%%

%%% Insert here the abstract %%%

\Abstract

This paper explores Neural Operators to predict turbulent flows, focusing on the Fourier Neural Operator (FNO) model. It aims to develop reduced-order/surrogate models for turbulent flow simulations using Machine Learning. Different model configurations are analyzed, with U-NET structures (UNO and U-FNET) performing better than the standard FNO in accuracy and stability. U-FNET excels in predicting turbulence at higher Reynolds numbers. Regularization terms, like gradient and stability losses, are essential for stable and accurate predictions. The study emphasizes the need for improved metrics for deep learning models in fluid flow prediction. Further research should focus on models handling complex flows and practical benchmarking metrics.\\

%%%  Insert here the actual article text %%% 

\section{Introduction} %%%%%%%%%%%%%%%%%%

Computational Fluid Dynamics is essential for studying turbulent flows and designing systems that interact with them. Simulations of turbulent flows at realistic scales require complex methods and a significant amount of computational power, making many of the high-fidelity simulations restricted to academic use and small space-time domains. A remedy to this is using Artificial Intelligence to enhance and accelerate CFD (\citet{brunton_machine_2020}, \citet{duraisamy_turbulence_2019}, \citet{vinuesa_potential_2021}). Current methods in Machine Learning can leverage significant volumes of data to create models that perform a wide array of tasks. In the context of CFD, according to \citet{vinuesa_potential_2021}, we can identify three main ways ML can contribute: turbulence modeling, accelerating DNS, and reduced-order modeling.\\

This work is centered on the development of reduced-order/surrogate models of turbulent flow simulations using Machine Learning.  This work aims to take a model based on the Fourier Neural Operator (\citet{li_fourier_2020}) and use it to predict turbulent flows over a long time horizon. We develop a training strategy for this type of model in order to achieve stable temporal predictions. Different model configurations are analyzed, and a training procedure is established in order to arrive at numerically stable and reasonably accurate predictions.\\

\section{Problem Statement: Learning to Simulate Turbulence from Data}

The subject of this research is the study of Deep Learning architectures for learning to replicate turbulent flow simulations. The goal for the Deep Learning model here is that given a set of initial conditions and/or other parameters, it can predict the next states of the turbulent flow as a numerical solver would do. We define the problem set as follows:

\begin{problem}
Being $u(\textbf{x}, t)$ the solution or family of solutions of a non-linear PDE in the form $\frac{du}{dt} = \mathcal{N}(u)$, where $\mathcal{N}$ is a non-linear operator, with boundary conditions $\textbf{x} \in \Omega \subset R^d$  and initial conditions $u(\textbf{x}, 0) = u_0$, where $u_0 \sim P(u_0)$. We want to find an approximation $\hat{u}$ of this solution discretized on spatio-temporal grid $\{\textbf{x}_i \in [0, L_i]\}$ and $\{ t_i \in [0, T) \}$ given a $\Delta x$ and $\Delta t$ respectively. This approximation is parameterized by a Neural Network $\hat{u} = f(\eta; \theta)$ with parameters $\theta$, that takes as input a set of features $\eta$, and whose parameters are found by solving the optimization problem $$ \theta = \underset{\theta}{\arg \min} \mathrm{E} \left[ L (u^{(i)}(\textbf{x}_i, t_i), \hat{u}^{(i)}(\textbf{x}_i, t_i),) \right] $$
\end{problem}

Where $L$ is a loss function that measures the difference between the data and its approximation, the ground-truth values $u^{(i)}(\textbf{x}_i, t_i)$ are drawn uniformly from a data set built from the results of a simulation. In the present study, we want to find a model that approximates solutions of the Navier-Stokes equation exhibiting turbulence. For doing so, we will remove the explicit time dependence of the Neural Network approximation by employing an auto-regressive model, where the time dependence is implicitly modeled because it takes as input features the previous state of the solution, and its output is the next stage of this one:

\begin{equation}
    \hat{u}_{t + \Delta t}(\textbf{x}) = f\left(\hat{u}_{t}(\textbf{x}); \theta \right)
\end{equation}

In a supervised learning setting, we find the parameters that minimize the mean-squared error between the predicted outputs and their target value. Given a data-set of $N$ samples composed of discrete trajectories of $u$ of length $T$ with a given $\Delta t$, the constrained risk minimization objective is:

\begin{equation}
\begin{gathered}
    \theta = \underset{\theta}{\arg\min} \sum_{i=1}^N \sum_{t_j=1}^{T-1} \left\| u^{(i)}_{t_{j+1}} - f(u^{(i)}_{t_j}; \theta) \right\| \\+ \mathbf{\lambda} \cdot L_{constraints} 
\end{gathered}
\end{equation}

Where $\mathbf{\lambda} \cdot L_{constraints}$ is a term for enforcing physical, statistical, or stability properties to the model predictions, as we will see later.

\section{Dataset: 2D Kolmogorov Flow}

The Kolmogorov flow (\citet{chandler_invariant_2013}) comes from solving the Navier-Stokes equation with a sinusoidal forcing:

\begin{equation}
\begin{gathered}
\frac{\partial \boldsymbol{u}^*}{\partial t^*}+\boldsymbol{u}^* \cdot \boldsymbol{\nabla}^* \boldsymbol{u}^*+\frac{1}{\rho} \boldsymbol{\nabla}^* p^* \\ =\nu \nabla^{* 2} \boldsymbol{u}^*+\chi \sin \left(2 \pi n y^* / L_y\right) \hat{\boldsymbol{x}} \\
\boldsymbol{\nabla}^* \cdot \boldsymbol{u}^*=0
\end{gathered}
\end{equation}

Where $\rho$ is the density, $\eta$ the kinematic viscosity, $n$ an integer describing the scale of the Kolmogorov force, and $\chi$ is the forcing amplitude per unit mass of fluid over a doubly periodic domain $\left[0, L_x\right] \times\left[0, L_y\right]$, $*$ indicates a dimensional quantity. The system is non-dimensionalized by the length-scale $L_y/{2\pi}$ and time scale $\sqrt{L_y/{2\pi \chi}}$, then the equations become:

\begin{equation}
\begin{gathered}
\frac{\partial \boldsymbol{u}}{\partial t}+\boldsymbol{u} \cdot \boldsymbol{\nabla} \boldsymbol{u}+\boldsymbol{\nabla} p=\frac{1}{R e} \nabla^2 \boldsymbol{u}+\sin (n y) \hat{\boldsymbol{x}}, \\
\boldsymbol{\nabla} \cdot \boldsymbol{u}=0
\end{gathered}
\end{equation}

Where the Reynolds number is:

\begin{equation}
R e:=\frac{\sqrt{\chi}}{v}\left(\frac{L_y}{2 \pi}\right)^{3 / 2}
\end{equation}

This equation is solved using doubly periodic boundary conditions in the velocity-vorticity formulation, which is obtained by taking the curl to the momentum equation:

\begin{equation}
\frac{\partial \omega}{\partial t}=\hat{z} \cdot \nabla \times(\boldsymbol{u} \times \omega \hat{z})+\frac{1}{R e} \nabla^2 \omega-n \cos (n y)
\label{eq:vor}
\end{equation}

Where $\omega \hat{z}:=\nabla \times \boldsymbol{u}$. This term is reduced to $-\boldsymbol{u} \cdot \nabla \omega$ since the vortex stretching is null in two dimensions ($\omega \cdot \nabla \boldsymbol{u}=0$). To solve this equation in two dimensions, the stream function Poisson equation is used to link the velocity to vorticity:

\begin{equation}
    \begin{aligned}
    & \boldsymbol{u} =\boldsymbol{\nabla} \times \psi(x, y) \hat{\boldsymbol{z}} \\
    & -\nabla^2 \psi = \omega
    \label{eq:stream}
\end{aligned}
\end{equation}

Equations (\ref{eq:vor}) and (\ref{eq:stream}) constitute the Stream-vorticity formulation of NS, and solving these yields an incompressible 2D flow.\\

For building the dataset, the equation is solved using a pseudo-spectral method implemented in Python using the pseudospectral module of the JAX-CFD package (\citet{bradbury_jax_2018}, \citet{dresdner_learning_2022}), where the Poisson equation is solved to find the velocity field, then the vorticity is differentiated, and the non-linear term is computed in physical space after it is dealised, then time is advanced using a Crank-Nicholson scheme update. The data is generated on a $256 \times 256$ uniform $(0,2 \pi )^2$ grid and downsampled to $128 \times 128$ for learning, the forcing is set to $f(\mathbf{x,y}) = 4\cos{(4(y))}$. Each trajectory is initialized with a Gaussian Random field to represent the initial velocity, where the maximum velocity is set to $7 m/s$. To build the dataset, solutions are recorded at each $t = 1s$ after the flow becomes stationary, and two datasets are built  for different $Re$:

\begin{table}[!h]
    \centering
        \begin{tabular}{|p{1.5cm}|p{1.5cm}|p{1.5cm}|p{1.5cm}|}
        \hline
         $\nu$ (viscosity) & $N$ (number of trajectories) & $T$ (time-steps) & $Re$ (Reynolds' Number)\\
         \hline
         $10^{-2}$ & 1000 & 200 & 100\\
         $2 \times 10^{-3}$ & 1000 & 200 & 500\\
         \hline
    \end{tabular}
    \caption{Datasets for 2D Kolmogorov Flow.}
    \label{tab:2d_kf_dataset}
\end{table}

\section{Deep Learning Models}

The simulations that the neural networks will learn are 2D CFD simulations or 2D slices of simulations in 3D. The models are trained with a fixed time step $\Delta t$ and fixed resolution $N_x \times N_y$. As stated in Section 2, we want the model to learn how to predict future fluid flow states given only an initial condition. This problem can be classified in ML terms as a spatiotemporal regression problem. In this case, we have a dataset of solutions to the NS equation defined in $\mathbf{R}^2$ that evolves in time. In other words, given an initial condition $u(\textbf{x}, t_0)$, it is necessary to find a model parameterized by $\theta$ that models the solution operator of the NS equation. In many numerical methods, the prediction of the next time step depends on the solution at one or more previous time steps. This means the model should learn the conditional probability:

\begin{equation}
    P_\theta(u) = \overset{N_k}{\underset{t=0}{\prod}}P_\theta(u_t | u_0, u_1, ..., u_{t-1})
\label{eq:ar_model}
\end{equation}

To be used in downstream applications, the learned model should:

\begin{itemize}
    \item Extrapolate to time horizons that are not present in the dataset.
    \item Retain the same degree of physical accuracy for all the predictions.
    \item Generalize to unseen initial conditions.
\end{itemize}

Many models can be used to model the solution of a PDE. However, obtaining a model that performs with these characteristics for turbulent flow data is challenging. Recurrent Neural Networks (RNN) are initially discarded because they are generally unstable to train due to vanishing/exploding gradients, and they struggle to learn long-range dependencies (\citet{goodfellow_deep_2016}, \citet{yousif_physics-guided_2022}). Physics-Informed Neural Networks are discarded too, even though they can learn a discretization invariant representation of the solution of a time-dependent PDE, they also present learning difficulties: they usually learn a solution of a PDE for a given set of initial and boundary conditions (\citet{raissi_hidden_2018}), they present problems in representing multi-scale phenomenon such as turbulence (\citet{wang_eigenvector_2020}, \citet{wang_understanding_2020}), and it can be difficult to generalize to log time-horizons (\citet{krishnapriyan_characterizing_2021}). Another critical aspect of the learned models is their data efficiency. RNNs and PINNs lack inductive biases. While PINNs achieve this through the PDE constraints, it results in an objective function that is challenging to optimize. To exploit the inductive bias of the data and to be able to predict solutions for an arbitrary duration, the models selected for this study are Autoregressive models based on Convolutions and Neural Operators. \\

The neural network architectures used in this study are the following:

\begin{itemize}
    \item \textbf{Fourier Neural Operator (FNO)}: Following their original implementation, the FNO (see figure \ref{fig:FNO_layer}) used for learning problems has 4 layers, each with 24 Fourier modes. The lifting layer is a fully connected layer with 32 features. The exact number of features is kept along all layers. The Swish activation function (\citet{ramachandran_searching_2017}) is used. This network's total number of parameters is \textbf{9.4M}.

    \begin{figure}[!h]
    \centering
    \includegraphics[width=\linewidth]{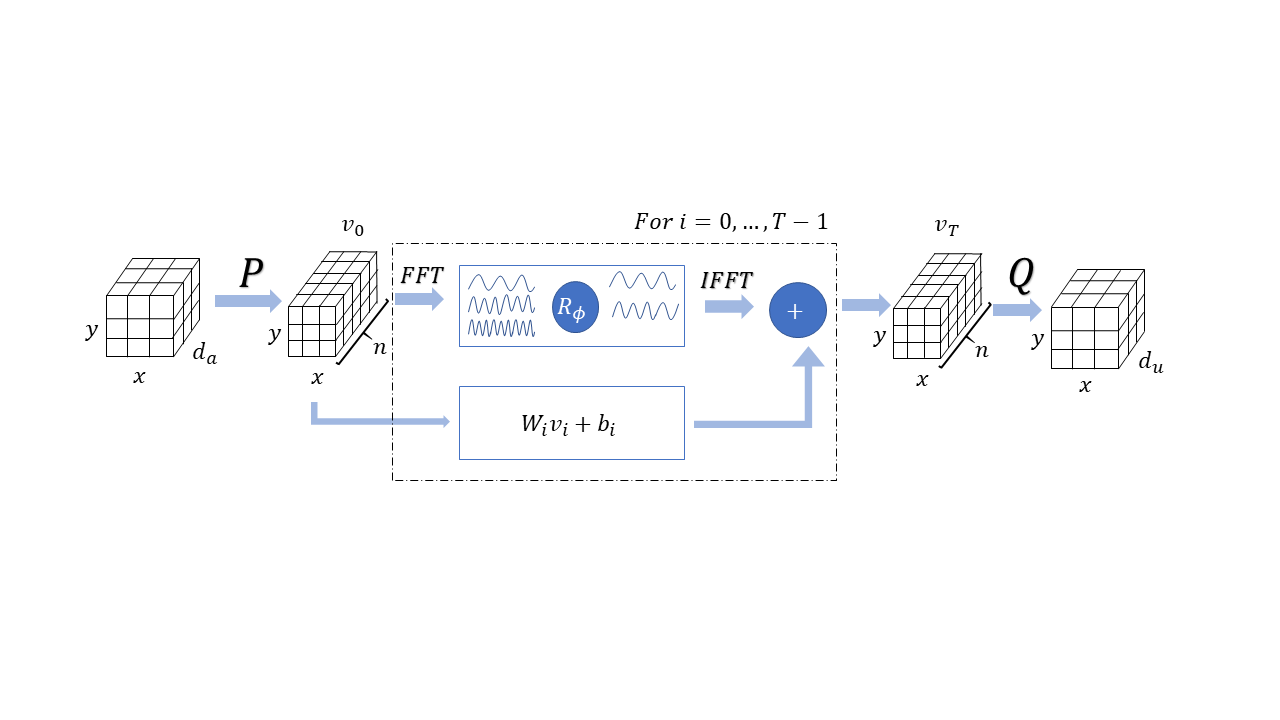}
    \caption{Diagram of an FNO architecture. The input tensor is lifted to a higher dimensional feature space. The FNO layer transforms its input to the Fourier domain, where it is filtered and then transformed back to the original domain. This operation is repeated $T$ times before passing to the projection layer that gives the output the dimensionality of the target function space.}
    \label{fig:FNO_layer}
    \end{figure}
    
    \item \textbf{U-shaped Neural Operator (UNO)}: The UNO model (\citet{rahman_u-no_2022}) as seen in figure \ref{fig:UNO} used has 3 encoder and decoder blocks and 1 processor block. The lifting and projection layers are Fully connected layers with 32 output features. The output layer is a linear layer that outputs to the target variable corresponding number of features. The encoding path reduces the spatial dimension by $3/4$ and increases the number of features by $3/2$, and the inverse applies to the decoding path. This means that we have three FNO layers for the decoder with $[ 48, 72,  108]$ features, respectively. The processor layer is just one FNO layer with no dimension or feature scaling. The output of each encoder block is skip-connected via concatenation to each decoder block. Every FNO layer in this model has a $k_{max} = 12$ mode. The Gaussian error Linear unit (GELU) (\citet{hendrycks_gaussian_2020}) is used as activation, and Layer Normalization (\citet{ba_layer_2016}) is used. This model's total number of parameters is \textbf{28M}. 

    \begin{figure}[!h]
    \centering
    \includegraphics[width=\linewidth]{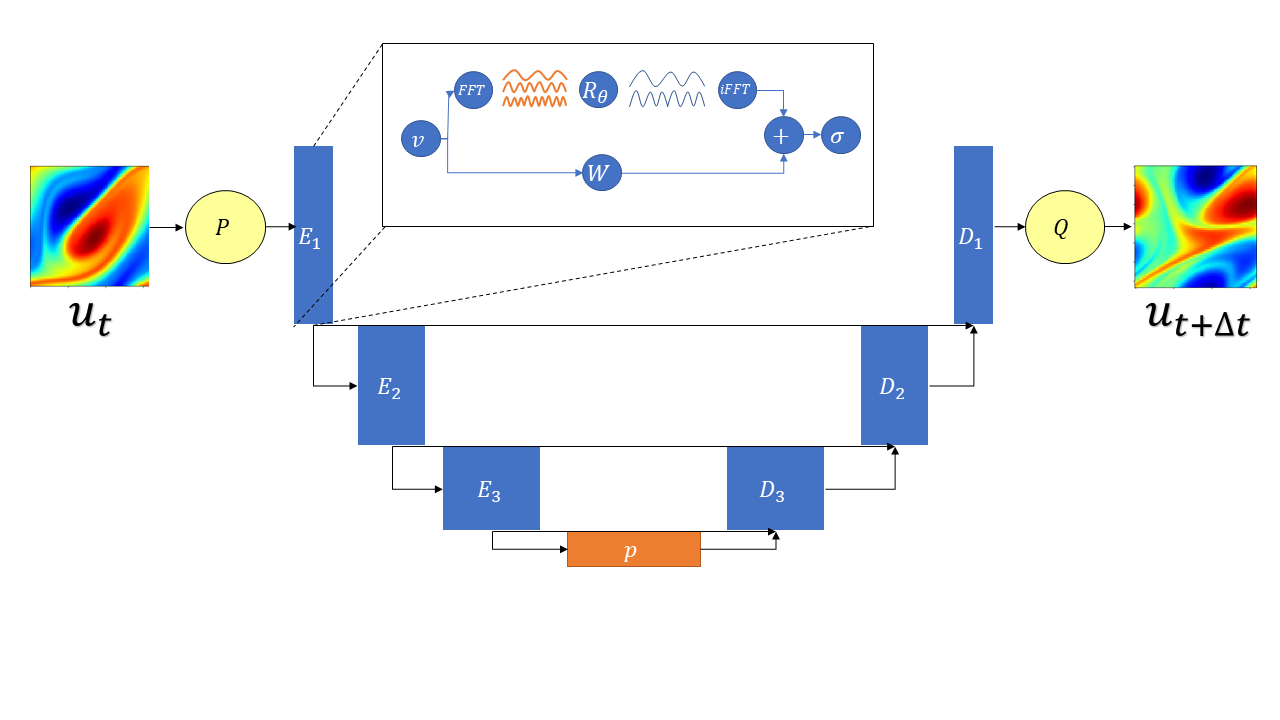}
    \caption{Diagram of the UNO architecture. Each block of the decoder is an FNO whose output is in a domain smaller than the previous. The inverse is the case for the decoder with the addition of a skip connection that connects each decoder block to the activations of corresponding encoder blocks. The middle processor layer is a standard FNO layer on the reduced domain.}
    \label{fig:UNO}
    \end{figure}
    
    \item \textbf{Fourier U-NET (U-FNET)}: This network follows a encoder-processor-decoder framework (\citet{sanchez-gonzalez_learning_2021},\citet{stachenfeld_learned_2022}). This standard U-NET model has intermediate FNO layers between the encoder and decoder. The encoder and decoder blocks are built with residual blocks (see fig. \ref{fig:unet_blocks}) with two convolutional layers of a kernel of size $3 \times 3$, one of the convolutions is zero-initialized,  GroupNorm (\citet{wu_group_2018}) with 1 group, and the GELU activation is used. The down and upsampling are performed using bilinear interpolation. Each block scales the dimensions by a factor of 2. The processor layer is an FNO with 3 layers and $k_{max}=8$ modes. The output layers are a $3\times 3$ convolutional layer followed by a linear, fully connected layer. The encoding path multiplies the features by $[2, 4, 16]$ respectively. The model designed this way has \textbf{25.9M} parameters.

\begin{figure}[!h]
    \centering
    \includegraphics[width=\linewidth]{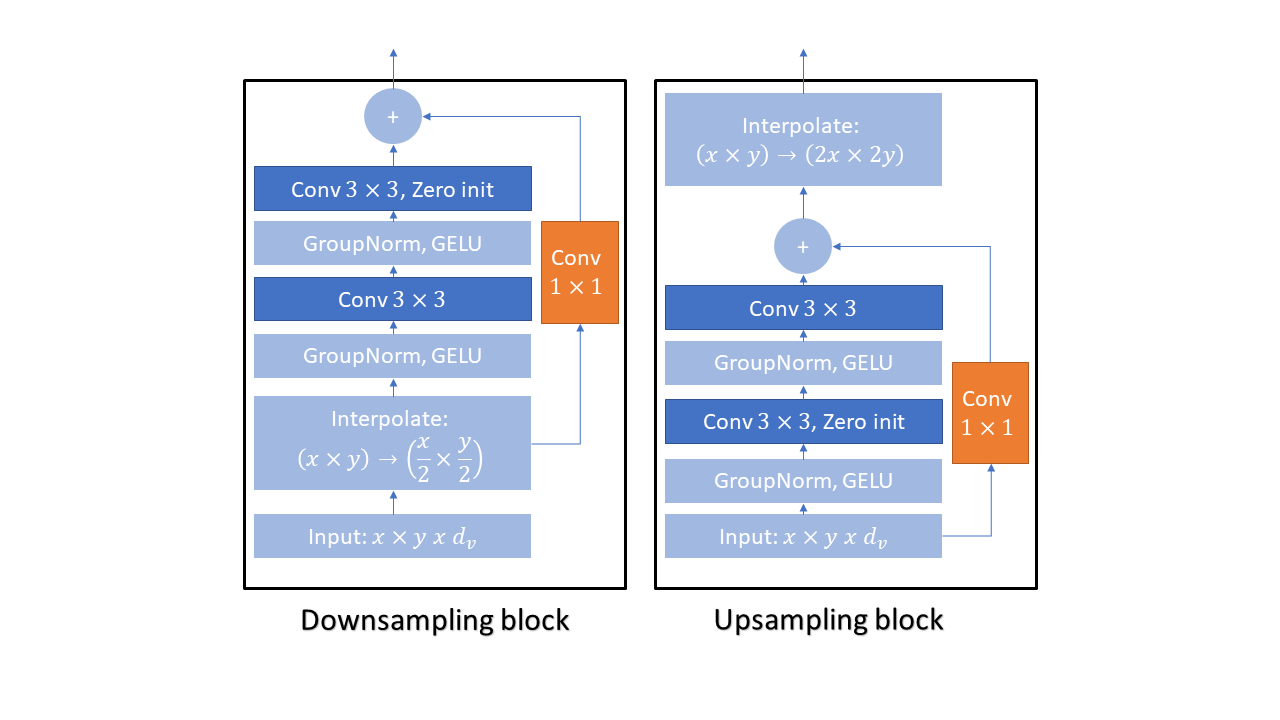}
    \caption{Downsampling and upsampling blocks for the U-FNET architecture.}
    \label{fig:unet_blocks}
\end{figure}

    \item \textbf{U-FNET-Euler}: This variation is the U-FNET model where the training target is to directly predict the difference between time steps instead of the following state. In previous work, it has been demonstrated that in some cases, learning the update instead of the complete state results in more accessible training \cite{sanchez-gonzalez_learning_2021}\cite{stachenfeld_learned_2022}. Also, there is a link with numerical integrators, where the AR model can be seen as an Euler discretization of the dynamics:

    \begin{equation}
        \begin{split}
            &\frac{dU}{dt} = \mathcal{G}_\theta(U) \\
            &\frac{U^{t_i} - U^{t_{i-1}}}{\Delta t} = \mathcal{G}_\theta(U^{t_{i-1}}) \\
            &U^{t_{i}} = U^{t_{i-1}} +  \mathcal{G}_\theta(U^{t_{i-1}})
        \end{split}
    \end{equation}
    \end{itemize}

\section{Gradient Loss term}

The multi-scale behavior of turbulent flows is a significant characteristic wherein features across both high and low-frequency ranges hold importance. It should be noted that Neural Networks encounter difficulties when attempting to learn functions with high fluctuations.  However, if optimization is solely focused on the $L^2$ loss using first-order optimization methods, the learning process may predominantly capture large-scale behavior, while smaller scales may require more time or may not be learned at all, resulting in noise within this range (\citet{rahaman_spectral_2019}).\\

To mitigate this issue, an approach is proposed wherein derivative information is incorporated into the loss function. Including derivative information reduces the errors associated with smaller scales, leading to an improved physical approximation. Some researchers have referred to this approach as Sobolev training, as described in the work by (\citet{czarnecki_sobolev_2017}). It is worth noting that this methodology shares similarities with physics-informed loss, as presented by (\citet{raissi_physics-informed_2019}). However, in Sobolev training, the focus is on the partial derivatives and relaxing the constraint on the partial differential equations (PDEs) while operating within the supervised learning framework.\\

For the construction of the Sobolev loss, we consider temporal and spatial derivatives up to the second order, as these derivatives are inherent in the Navier-Stokes equations:

\begin{equation}
    \begin{split}
        L_{pde} = \mathbb{E} \left [\left \|\nabla\hat{U} - \nabla U 
    + \nabla^2\hat{U} - \nabla^2 U + \frac{dU}{dt} - \frac{d\hat{U}}{dt} \right \|^2 \right]
    \end{split}
\end{equation}

\section{Promoting stability through regularization}

Turbulent flows are characterized by their chaotic nature, rendering them challenging to predict due to their high sensitivity to perturbations. Even a minor deviation from the expected trajectory can result in a significantly different future path, commonly called the butterfly effect. The prediction of chaotic systems' evolution poses difficulties for numerical methods due to the rapid propagation of small numerical errors, leading to non-physical outcomes. To address this issue, numerical schemes typically employ robust time-integration techniques and minimize the time step ($\Delta t$) to maintain accuracy. On the other hand, numerical methods achieve numerical stability by their numerical dissipation properties that can be inherent to the method or enforced.\\

Similarly, Neural Networks encounter obstacles when attempting to predict chaotic behavior. Previous studies have utilized Recurrent Neural Networks (RNNs) for predicting attractors in dynamical systems with finite-dimensional state spaces or lower complexity Partial Differential Equations (PDEs) (\citet{lesjak_chaotic_2021}, \citet{gilpin_chaos_2021}, \citet{gelbrecht_neural_2021}). However, since understanding neural networks' numerical properties is limited, it is unsure whether a NN can exhibit numerical dissipation or be numerically stable for arbitrary time-series data.\\

Temporal stability has been a critical concern in training Neural Networks for predicting solutions to the Navier-Stokes and Euler equations, as highlighted in other research papers (\citet{sanchez-gonzalez_learning_2020}, \citet{brandstetter_message_2022}, \citet{li_learning_2022}). What these papers do essentially is to try to learn dissipative dynamics through regularization. This is similar to training for denoising. The inputs are perturbed by adversarial attacks coming from a different probability distribution. Then, the model is taught to mitigate or suppress this noise by minimizing the distance of the prediction with the added perturbation $\epsilon$ and the target state. A stability term is introduced into the loss function following the formulation:
\begin{equation}
\begin{aligned}
    L_{stability} &= \mathbb{E}\left[\mathcal{L}(\mathcal{G}_\theta(U^k+\epsilon), U^{k+1})\right] \\
    L_{stability} &= \frac{1}{N_{data}}\frac{1}{N_t}  \overset{N_{data}}{\underset{n = 1}{\sum}} \overset{N_t}{\underset{i = 1}{\sum}} \left\|U^{t_i}_n - \mathcal{G}_\theta(U^{t_{i-1}}_n + \epsilon) \right\|^2
\end{aligned}
\end{equation}

\begin{figure}[!h]
    \centering
    \includegraphics[width = \linewidth]{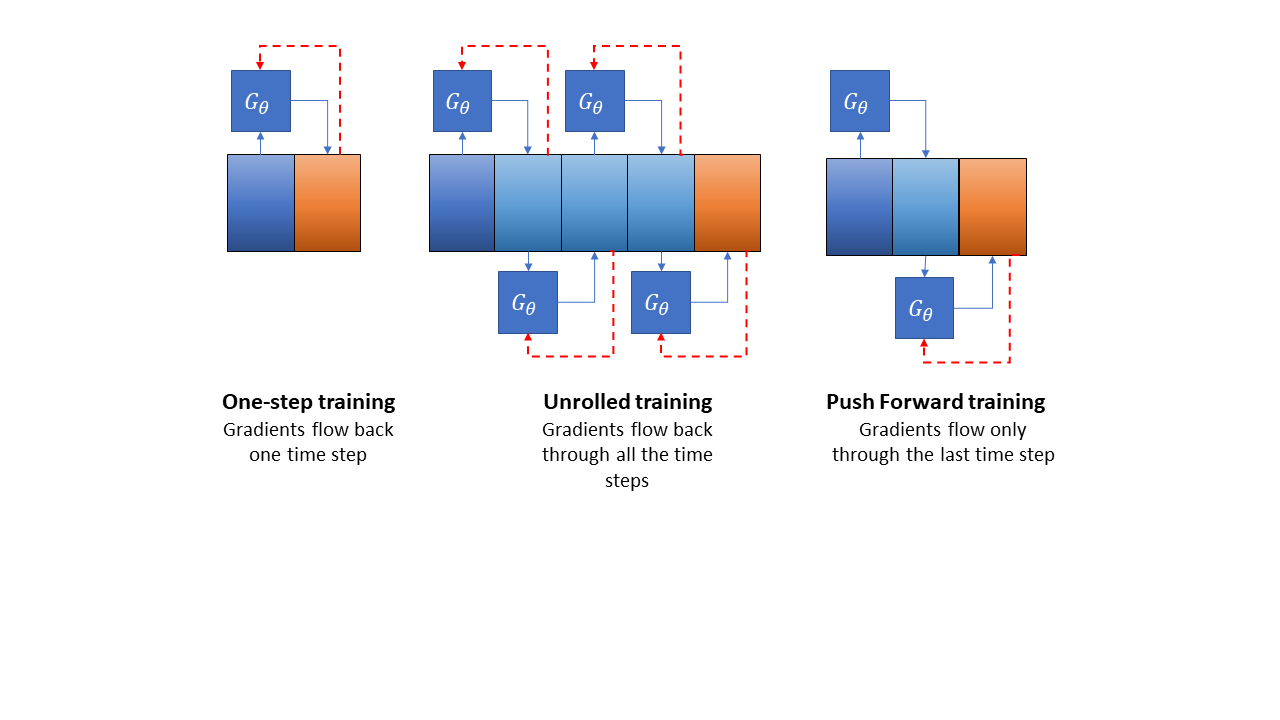}
    \caption{Training strategies. One-step training predicts only the next time step. Unrolled training predicts a whole sequence given only an initial time step and backpropagates (red line) through it. The model is unrolled $N$ times in pushforward training but only backpropagates to the last time step.}
    \label{fig:pushf}
\end{figure}

The next step consists of determining how to add the perturbations to the inputs of our model. There are mainly two approaches that have been used to promote stability in Autoregressive Neural Networks. The first method is adding Gaussian noise to the inputs with constant variance (\citet{sanchez-gonzalez_learning_2020}). This study uses Gaussian Noise of $\sigma = 0.01$. The other method used to add the perturbations is to let the model do unrolled predictions for a small amount of the time steps and then only backpropagate to the last time step. This is called the pushforward method (\citet{brandstetter_message_2022}), (see fig. \ref{fig:pushf}).

This work will use pushforward and denoising training methods to promote stability. First, models are trained with the pushforward method. When the results converge in terms of the training MSE, the stability loss is switched with the denoising method to help enforce numerical dissipation further. In this work, combining both techniques bears a more accurate result than exclusively using one of these two losses. \\

\section{Results and discussion}

In this section, the performance of the models is evaluated on the 2D Kolmogorov Flow. First, the errors for a 1-step prediction of the model can be observed. Table \ref{tab:errors_2dkf} presents the RRMSE and the PDE loss values calculated on test data not used during training. From this, it can be inferred that the models with the U-NET structure have better accuracy than the standard FNO. However, this measures only how close the model is to the ground truth, so it is not enough to assess the numerical stability and physical accuracy of the model.\\ 

\begin{table}[!h]
    \centering
    {\rowcolors{2}{blue!80!white!50}{white!70!blue!40}
    \begin{tabular}{|c|c|c|}
        \hline
         Model & RRMSE & $L_{pde}$ \\
         \hline
         FNO & 0.2771 &  0.0020\\
         UNO & 0.2966 & 0.0132 \\
         U-FNET-Euler & 0.1409 & 0.0010\\
         U-FNET & 0.19850 & 0.0138\\
         \hline
    \end{tabular}
    }
    \caption{Errors for each model for 1-step prediction of the 2D Kolmogorov flow.}
    \label{tab:errors_2dkf}
\end{table}

\begin{figure}[!h]
    \centering
    \includegraphics[width=\linewidth]{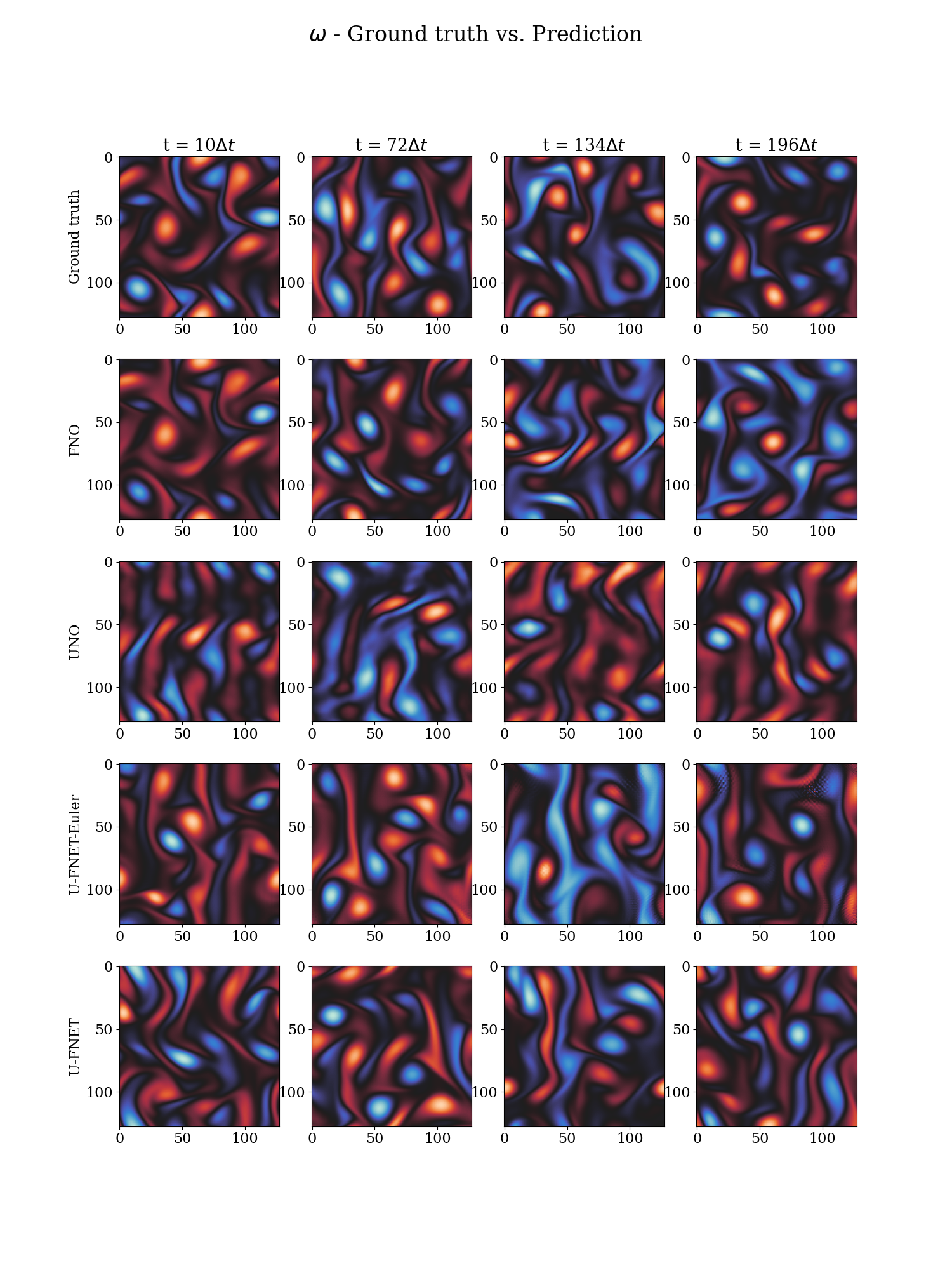}
    \caption{Samples of vorticity predictions by different models}
    \label{fig:samples}
\end{figure}

Colormaps of vorticity predicted by the models studied can be seen in figure \ref{fig:samples}. This serves as a qualitative assessment in the sense that the neural networks are able to sustain predictions with similar structures given the same initial condition. However, from this picture, it is observed too that even parting from the same initial state, the predictions of each neural network are different than those of the simulation. The question arises whether this can still be considered a solution to the Navier-Stokes equation.\\

In order to verify that the flow predicted by the models is turbulent, we assess statistical properties. The autocorrelation function provides a quantitative measure of how long it takes for the flow variable to lose its correlation with its past values, allowing researchers to identify dominant time scales and capture important flow features.  In a general sense, the autocorrelation is given by:\\

\begin{equation}
        R_{f f}(\tau)=\int_{-\infty}^{\infty} f(t+\tau) \overline{f(t)} \mathrm{d} t=\int_{-\infty}^{\infty} f(t) \overline{f(t-\tau)} \mathrm{d} t
\end{equation}

\begin{figure}[!h]
    \centering
    \includegraphics[width=\linewidth]{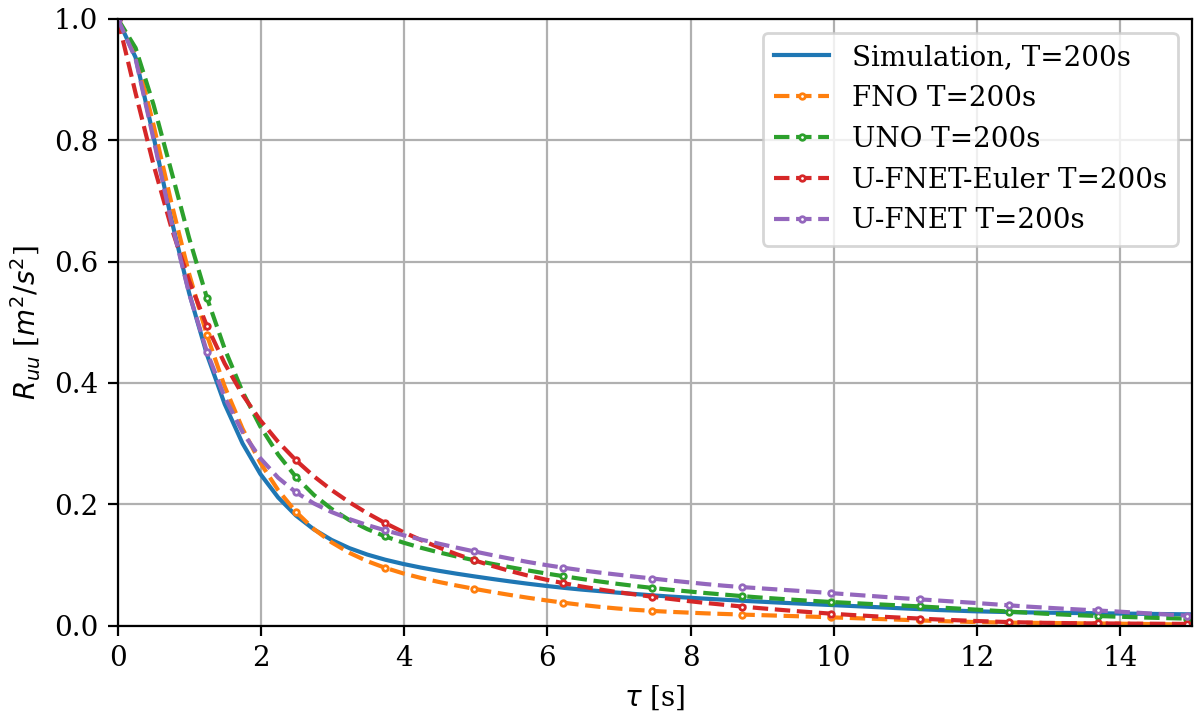}
    \caption{Autocorrelation function of vorticity}
    \label{fig:R}
\end{figure}

Where $\tau$ is the time-lag, and $\overline{f}$ represents the complex conjugate of the signal that, in the cases studied, is the same signal since it is real. The Autocorrelation of the vorticity fluctuations is observed in figure \ref{fig:R}. \\

The integral of the Autocorrelation function indicates the turbulent time-scale. This is an invariant property independent of the flow realizations. Table \ref{tab:tl} presents the integral time-scale value for each model and their relative mean absolute error w.r.t. the $\tau_l$ of the simulation.

\begin{table}[!h]
    \centering
    {\rowcolors{2}{blue!80!white!50}{white!70!blue!40}
    \begin{tabular}{|c|c|c|}
        \hline
         Model & $\tau_l (s) $ & r-MAE \\
         \hline
         Sim &  2.15 & -- \\
         FNO & 1.75 s &  0.185\\
         UNO & 2.31 & 0.073 \\
         U-FNET-Euler & 2.07 & 0.038\\
         U-FNET & 2.13 & 0.012\\
         \hline
    \end{tabular}
    }
    \caption{Integral time-scale and their relative mean absolute error.}
    \label{tab:tl}
\end{table}

\begin{figure}[!h]
    \centering
    \includegraphics[width=\linewidth]{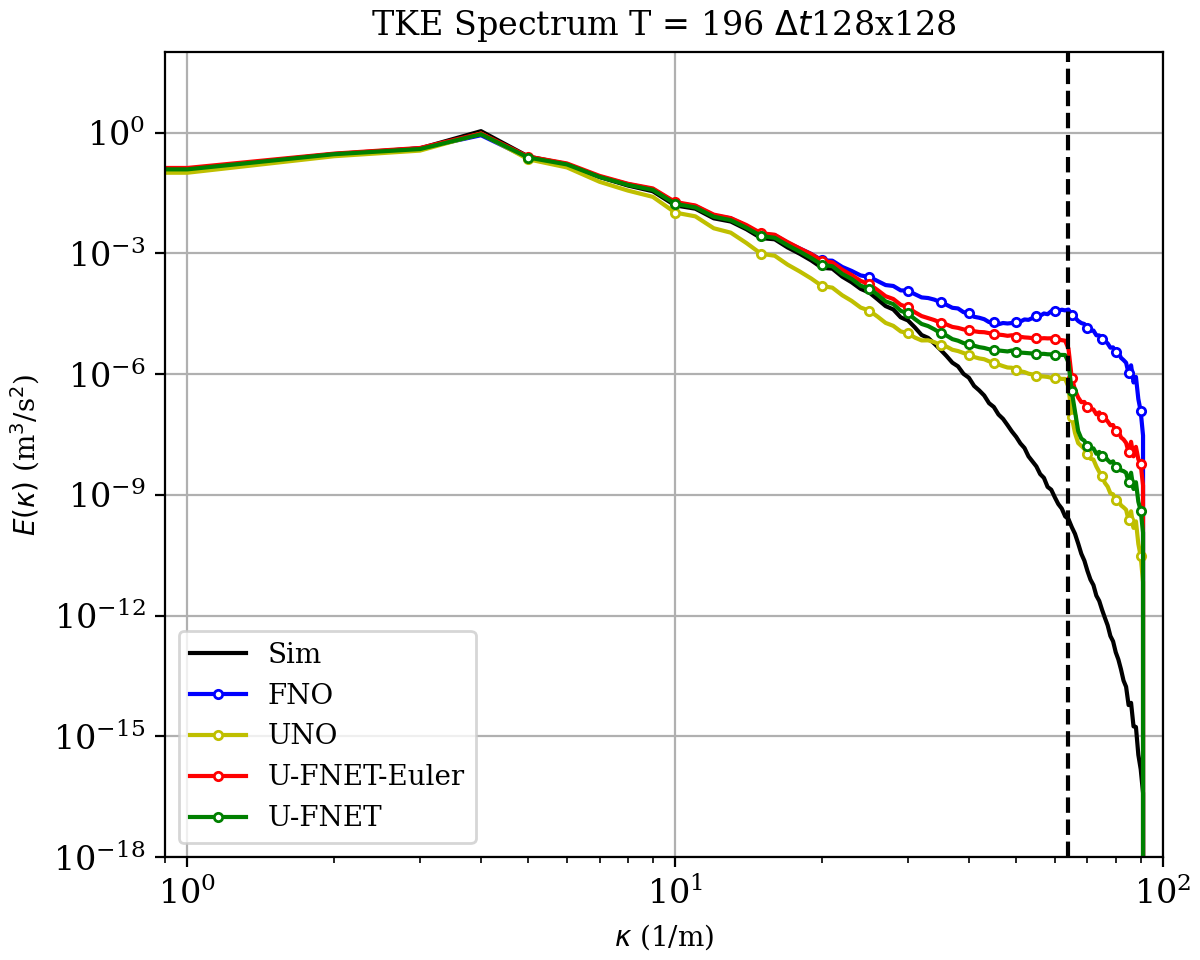}
    \caption{Energy spectrum at $T = 196\Delta t$.}
    \label{fig:tke}
\end{figure}

Next, the kinetic energy spectrum is reported for a time-step near the final time-step available in the dataset. As can be seen in figure \ref{fig:tke}, the models have a very low error on the high-energy-containing eddies, while the error in the smaller scales is higher.\\

The MAE in Fourier space is reported in the table \ref{tab:tke_er}. There are two errors, one for the lower frequencies and the other for the higher ones. The division is made near the point where the slope of the spectrum changes.\\

\begin{table}[!h]
    \centering
    {\rowcolors{2}{blue!80!white!50}{white!70!blue!40}
    \begin{tabular}{|c|c|c|}
        \hline
         Model & MAE-low & MAE-high \\
         \hline
         FNO & 0.0107 &  $1.20 \times 10^{-5}$\\
         UNO & 0.0132 &  $4.45 \times 10^{-7}$ \\
         U-FNET-Euler & 0.0058 &  $3.04 \times 10^{-6}$\\
         U-FNET & 0.0066 &  $1.24 \times 10^{-6}$\\
         \hline
    \end{tabular}
    }
    \caption{TKE error for the different models.}
    \label{tab:tke_er}
\end{table}

In order to determine which model is better, by looking at the results presented, we can conclude that the best ones are the ones with the U-NET structure, UNO, and U-FNET. The U-FNET-Euler is discarded because it was not able to be stable after 200 time steps. On another hand, the standard FNO exhibited stability, but the errors in autocorrelation and kinetic energy spectrum were higher than the others.\\

\subsection{Results at a higher Reynolds number}

The next question to be addressed is which model is able to predict turbulence at a higher Reynolds number. In the experiments carried out, the only model that was able to converge was the U-FNET. The other models got stuck at a point where no turbulent fluctuations were produced. To train the U-FNET, transfer learning was used, meaning that the weights of the trained model at $Re = 100$ were used as the initial weights for training the model at $Re = 500$. The relative root mean squared error, and the gradient loss is seen in table \ref{tab:errors_re500}.\\

\begin{table}[!h]
    \centering
    {\rowcolors{2}{blue!80!white!50}{white!70!blue!40}
    \begin{tabular}{|c|c|c|}
        \hline
         Model & RRMSE & $L_{pde}$ \\
         \hline
         U-FNET & 0.1614 & 0.0343\\
         \hline
    \end{tabular}
    }
    \caption{Errors for U-FNET predictions at $Re = 500$.}
    \label{tab:errors_re500}
\end{table}

\begin{figure}[!h]
    \centering
    \includegraphics[width=\linewidth]{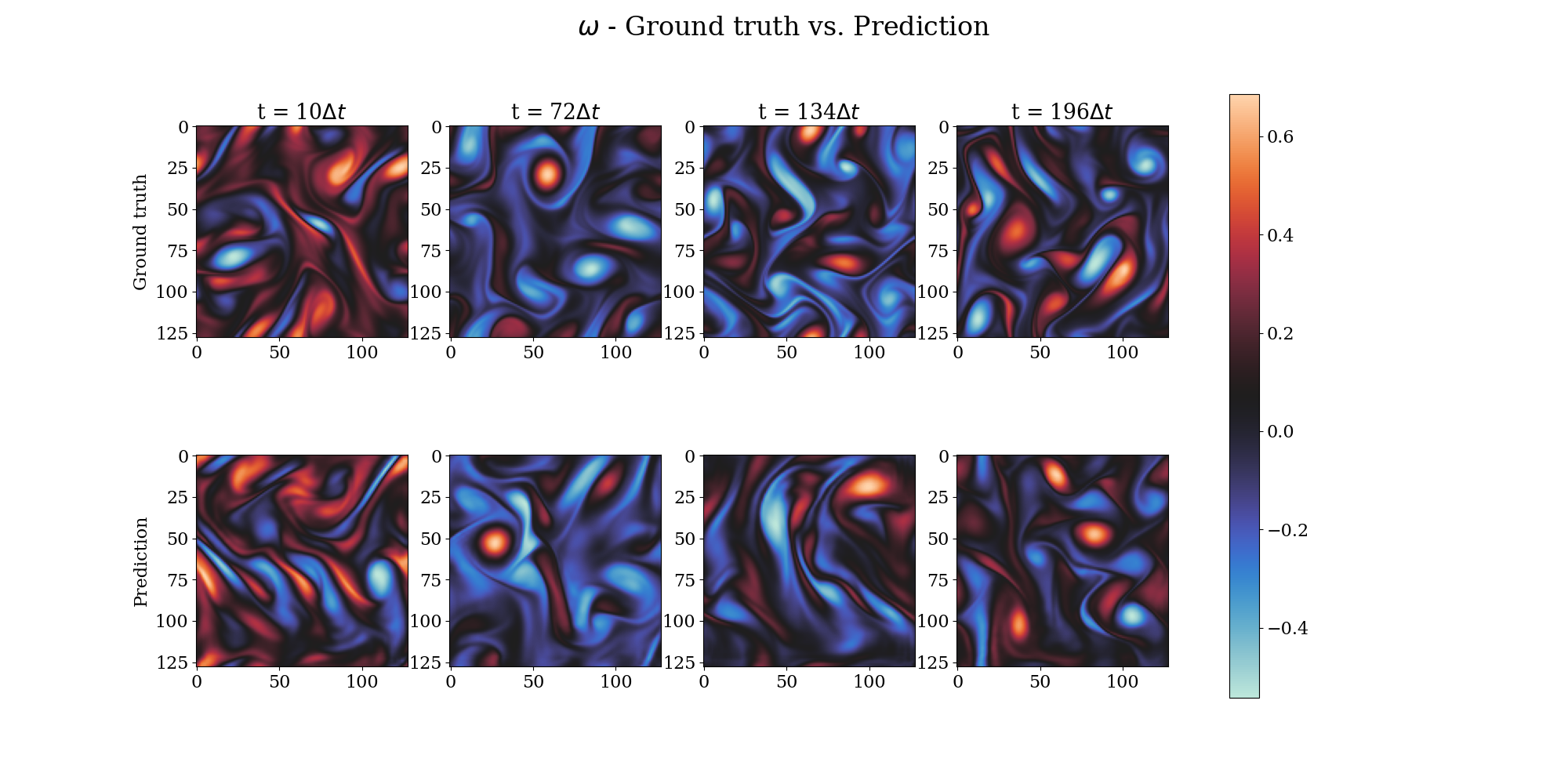}
    \caption{Samples of vorticity predictions by U-FNET at $Re = 500$}
    \label{fig:samples_500}
\end{figure}

\begin{figure}[!h]
    \centering
    \includegraphics[width=\linewidth]{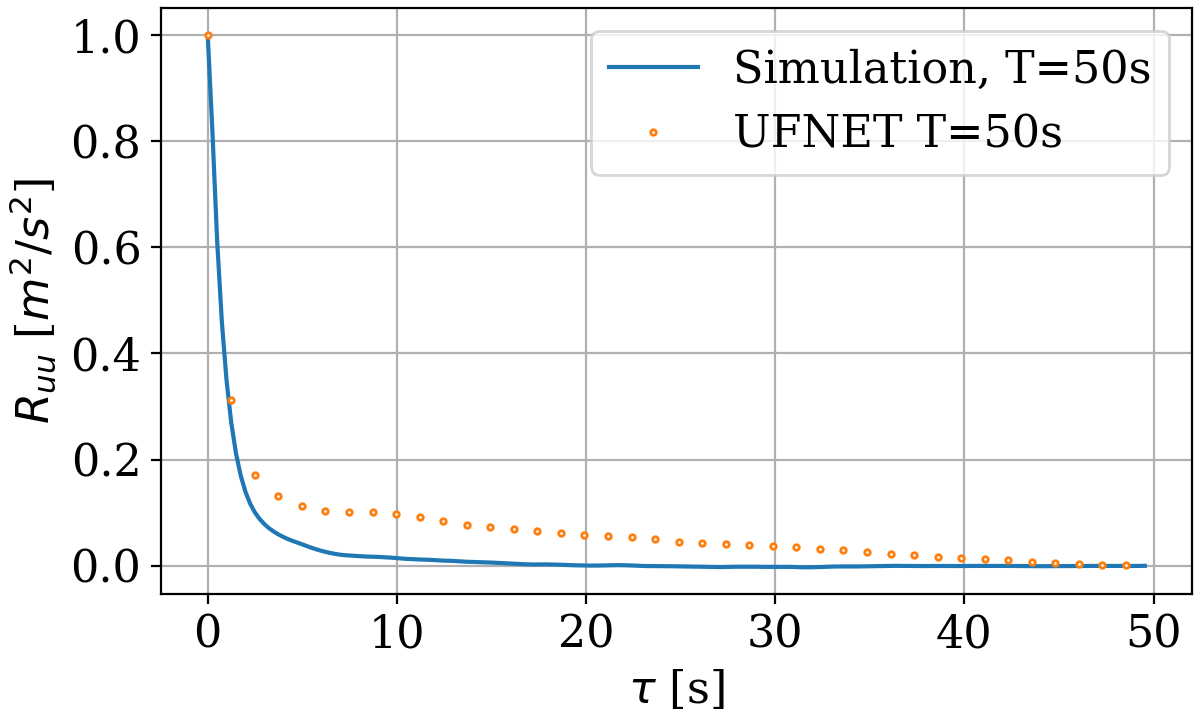}
    \caption{Autocorrelation function of vorticity at $Re = 500$}
    \label{fig:R_500}
\end{figure}

\begin{figure}[!h]
    \centering
    \includegraphics[width=\linewidth]{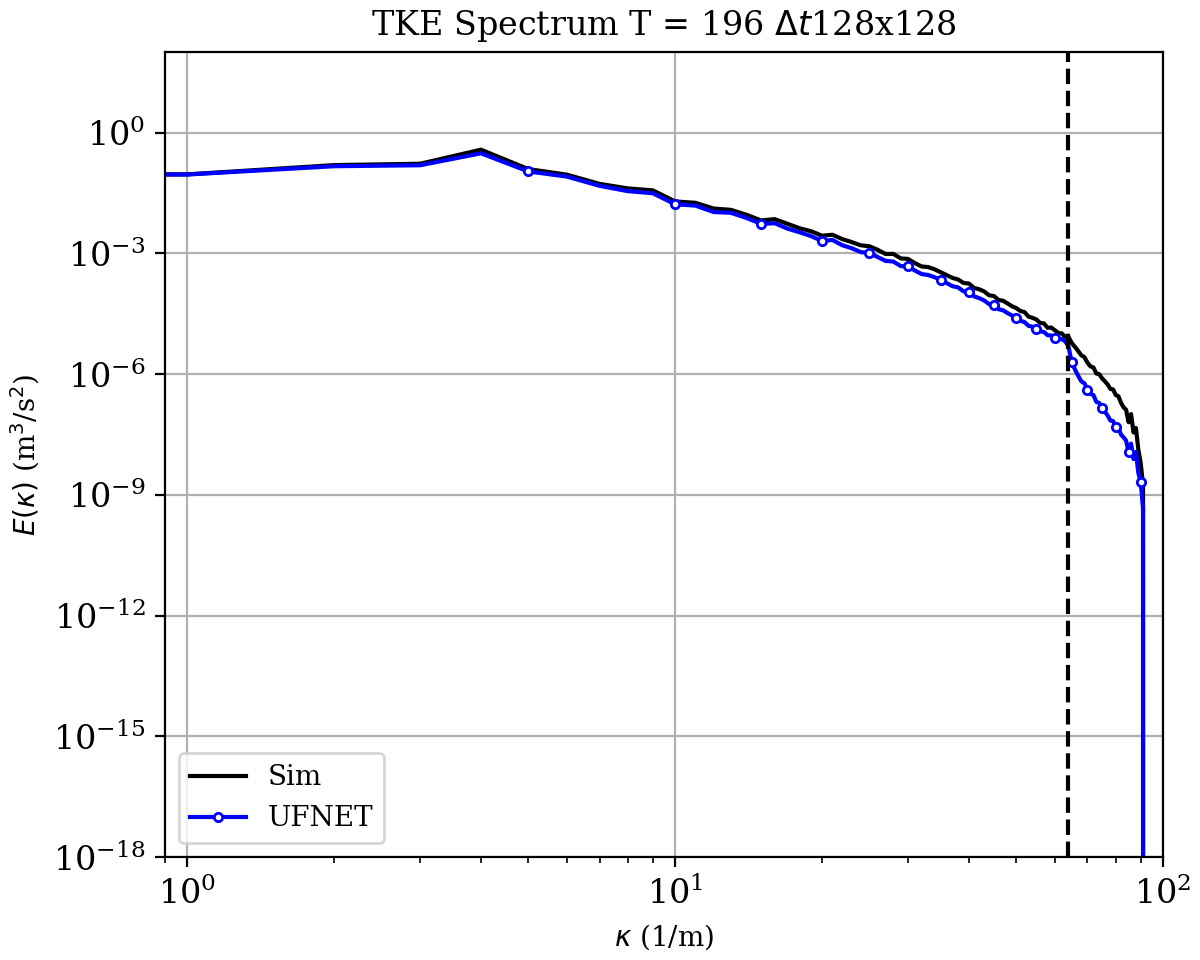}
    \caption{Energy spectrum at $Re = 500$.}
    \label{fig:tke_500}
\end{figure}

Nest, autocorrelation (fig. \ref{fig:R_500}), and TKE spectrum fig. \ref{fig:tke_500} are presented. There is good agreement with the simulation data, with errors close to the ones of the U-FNET at a lower $Re$ as it can be seen in tables \ref{tab:errors_re500}, \ref{tab:tl_500} and \ref{tab:errors_tke_re500}. There is no clear reason why only this model worked for this problem. Being the only difference in the convolutional layers, it can be concluded that this is possibly the reason because the convolutions learn representations based on local features that make learning in the Fourier space easier. This needs further investigation but is a sign that embedding inductive biases in the models can help learn complex features. \\

\begin{table}[!h]
    \centering
    {\rowcolors{2}{blue!80!white!50}{white!70!blue!40}
    \begin{tabular}{|c|c|c|}
        \hline
         Model & $\tau_l (s) $ & r-MAE \\
         \hline
         Sim &  1.23 & -- \\
         U-FNET & 1.22 & 0.010\\
         \hline
    \end{tabular}
    }
    \caption{Integral time-scale and their relative mean absolute error.}
    \label{tab:tl_500}
\end{table}

\begin{table}[!h]
    \centering
    {\rowcolors{2}{blue!80!white!50}{white!70!blue!40}
    \begin{tabular}{|c|c|c|}
        \hline
         Model & MAE-{$TKE$}-low &  MAE-{$TKE$}-high\\
         \hline
         U-FNET & 0.0048 & $1.09 \times 10^{-5}$\\
         \hline
    \end{tabular}
    }
    \caption{Errors for U-FNET predictions at $Re = 500$.}
    \label{tab:errors_tke_re500}
\end{table}

\subsection{Effect of the regularization terms in learning}

\begin{figure}[!h]
    \centering
    \includegraphics[width=\linewidth]{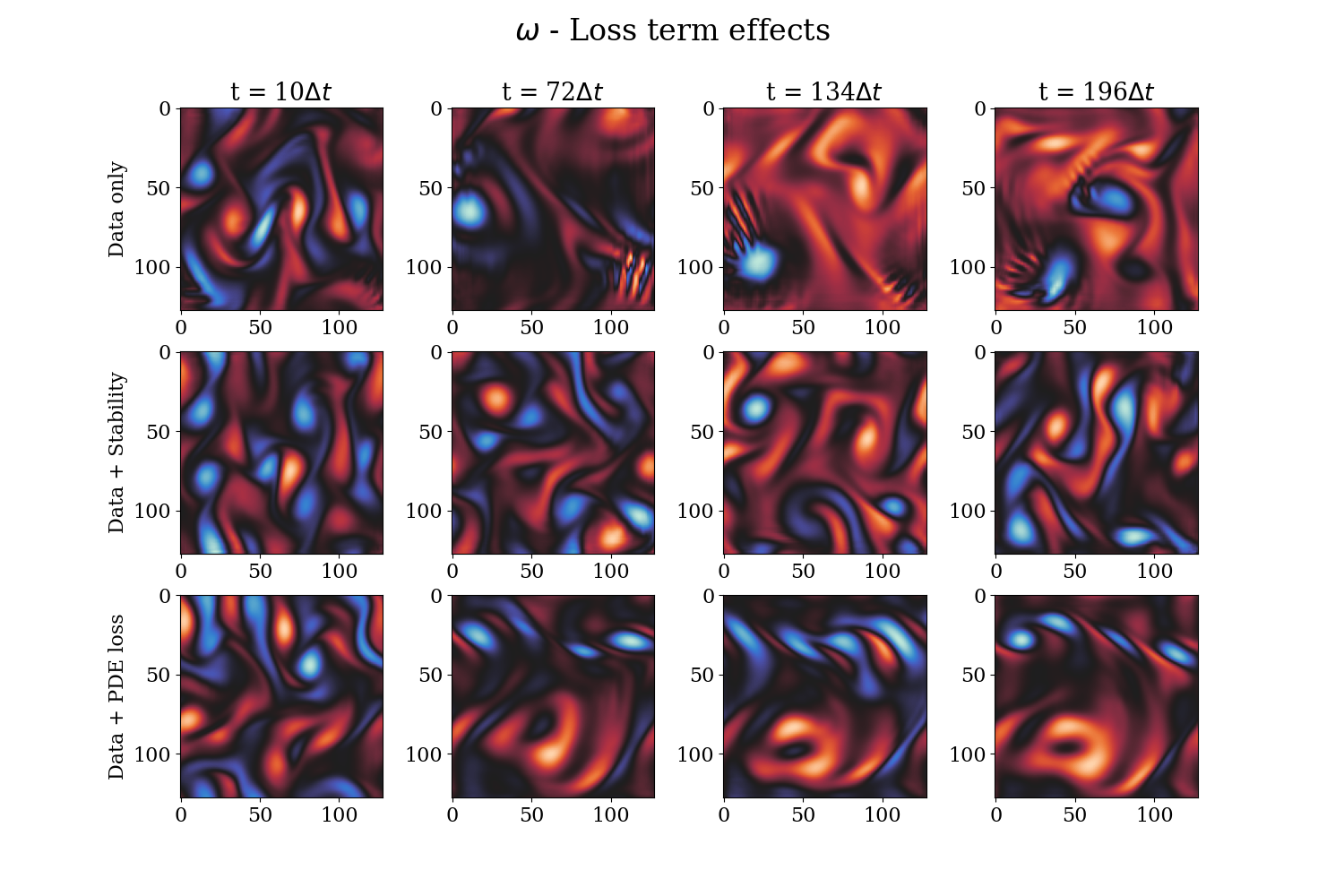}
    \caption{Samples of U-FNET trained with different terms in the loss function}
    \label{fig:samples_effects}
\end{figure}

To determine the role the terms of the loss functions play in learning, a study is performed where the U-FNET model is trained only with the MSE, other with the MSE and the gradient loss terms and other with the MSE with the stability term. Figure \ref{fig:samples_effects} shows samples of the U-FNET trained with the different losses. According to the results, the model trained with the stability term in it is the only one producing coherent results. According to table \ref{tab:errors_effects}, the model trained with the stability term has higher errors than the other models. However, this is the only one producing good predictions. This is an important indication that distance-based metrics are insufficient to quantify the performance of deep learning models that predict fluid flow.\\

\begin{table}[!h]
    \centering
    {\rowcolors{2}{blue!80!white!50}{white!70!blue!40}
    \begin{tabular}{|c|c|c|}
        \hline
         Model & RRMSE & $L_{pde}$ \\
         \hline
         Data-only & 0.2048 &  0.0269\\
         Data+PDE & 0.1819 & 0.0179 \\
         Data+Stability & 0.2204 & 0.0296\\
         \hline
    \end{tabular}
    }
    \caption{Errors for each model for 1-step prediction of the 2D Kolmogorov flow.}
    \label{tab:errors_effects}
\end{table}

\begin{figure}[!h]
    \centering
    \includegraphics[width=\linewidth]{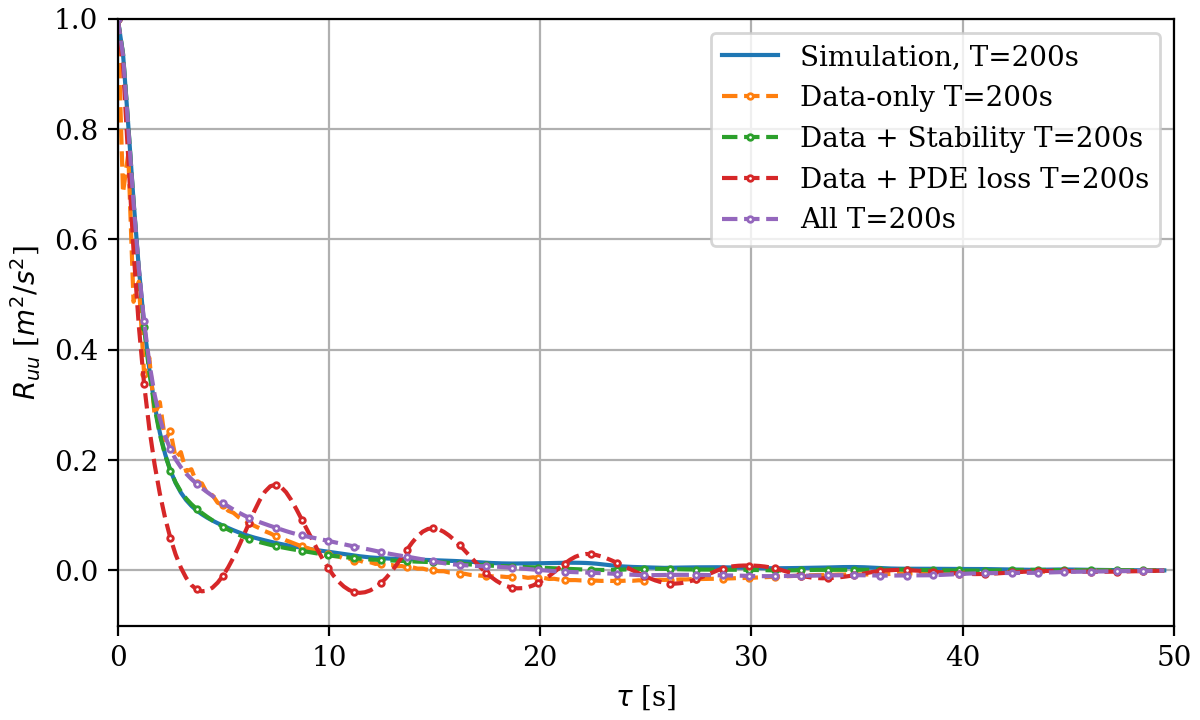}
    \caption{Autocorrelation function of U-FNET trained with different terms in the loss function}
    \label{fig:R_effects}
\end{figure}

The autocorrelation of the model is shown in Fig. \ref{fig:R_effects}. This figure confirms that the model trained with stability term has a temporal behavior closer to the ground truth data. This is quantified by the error in the integral time-scale reported in table \ref{tab:tl_effects}, with an error higher than the model trained with all the regularization on.\\

\begin{table}[!h]
    \centering
    {\rowcolors{2}{blue!80!white!50}{white!70!blue!40}
    \begin{tabular}{|c|c|c|}
        \hline
         Model & $\tau_l (s) $ & r-MAE \\
         \hline
         Sim &  2.15 & -- \\
         Data-only & 1.67 s &  0.185\\
         Data+PDE & 1.44 & 0.328 \\
         Data+Stability & 1.97 & 0.084\\
         All & 2.13 & 0.012\\
         \hline
    \end{tabular}
    }
    \caption{Integral time-scale and their relative mean absolute error.}
    \label{tab:tl_effects}
\end{table}

\begin{figure}[!h]
    \centering
    \includegraphics[width=\linewidth]{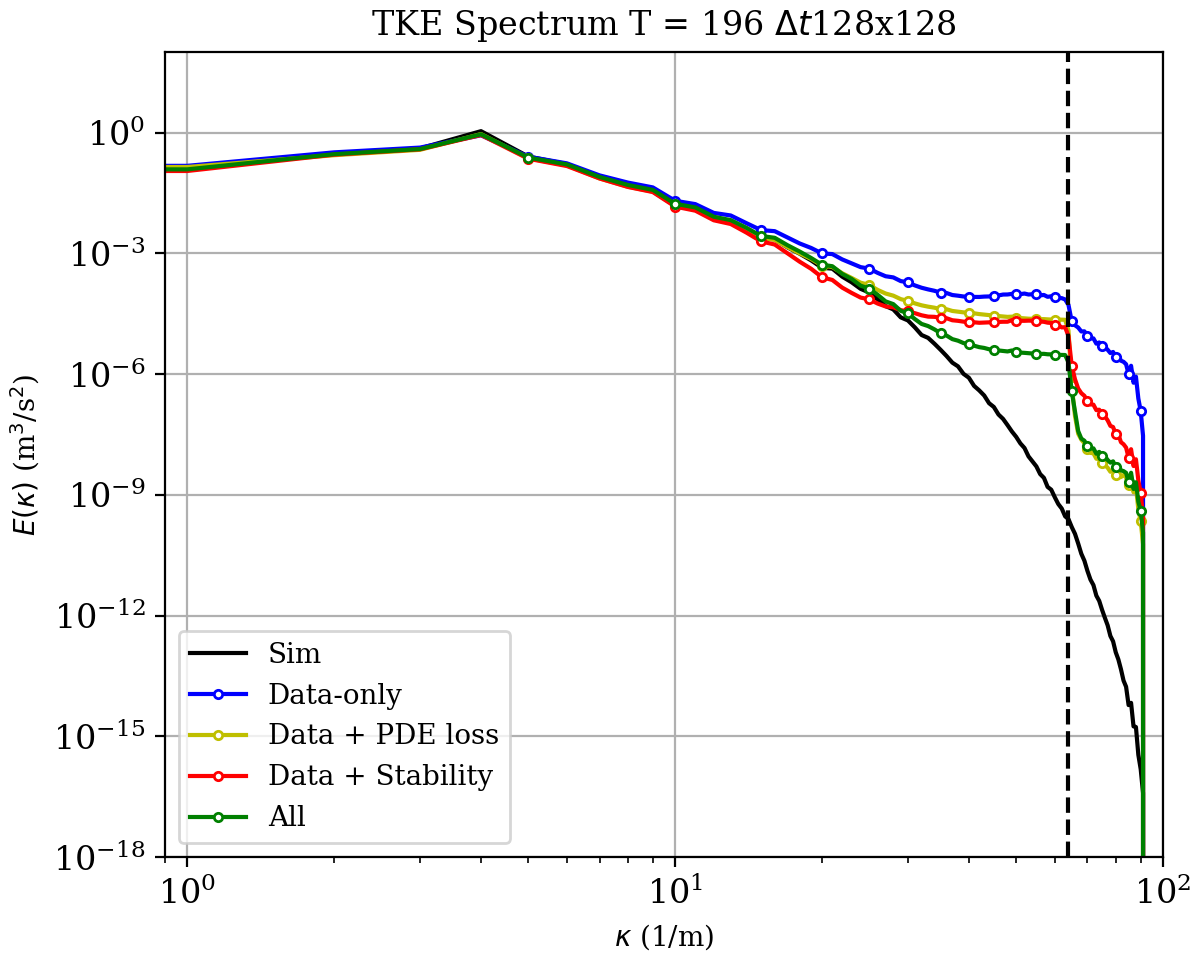}
    \caption{Energy spectrum of U-FNET trained with different terms in the loss function}
    \label{fig:tke_effect}
\end{figure}

The kinetic energy spectrum is shown in Fig. \ref{fig:tke_effect}. In this case, the model trained with the PDE term outperforms the one with stability in the lower frequencies as seen in table \ref{tab:tke_er_effect}. This is a case for overfitting since the model learns a solution that minimizes the terms that fit the spatial scales, but it does not optimize the temporal behavior, reinforcing the hypothesis of the need for a training procedure that enforces temporal coherence.\\

\begin{table}[!h]
    \centering
    {\rowcolors{2}{blue!80!white!50}{white!70!blue!40}
    \begin{tabular}{|c|c|c|}
        \hline
         Model & MAE-low & MAE-high \\
         \hline
         Data-only & 0.0115 &  $2.99 \times 10^{-5}$\\
         Data+PDE & 0.0083 &  $8.67 \times 10^{-6}$ \\
         Data+Stability & 0.0111 &  $6.19 \times 10^{-6}$\\
         All & 0.0066 & $1.24 \times 10^{-6}$\\
         \hline
    \end{tabular}
    }
    \caption{TKE error for the different models.}
    \label{tab:tke_er_effect}
\end{table}

\section{Conclusions}

We have presented the use of Neural Operators to predict turbulent flows over a long period of time. We studied configurations based on the Fourier Neural Operator, which is considered one of the state-of-the-art models in Operator Learning for physics and science problems. From the results presented, it can be concluded that training these models to predict spatiotemporal phenomena with chaotic behavior is not an easy task. Learning with data in a supervised learning setting is not enough, and strategies to enforce numerical stability and temporal coherence are needed. The study also found that using a mix of Convolutional and FNO layers can improve performance in these tasks and enable learning more complex behavior without increasing model size. This study opened more questions regarding the limitations of these models and ways to improve them. The flow configuration used for training is a relatively simple one. that is not found in nature. The use of these types of models on more complex physics like multiphase and turbulence in 3D has been very modest and would require different models and training strategies to tackle the higher dimensionality of these problems. There is a wide perspective regarding the use of machine learning in surrogate modeling of fluid flows that is beyond the scope of this article. Future work, we believe, should be oriented towards models and training methods that are able to tackle more complex flows with better generality to different parameters, as well as appropriate benchmarking metrics that quantify the trustworthiness of ML as a substitution of numerical methods to be used in practical cases. \\

%%% Insert here acknowledgments if necessary %%%

\Acknowledgments

This work was granted access to the HPC resources of IDRIS under the allocation 2023-AD011013548R1 made by GENCI. 

%%% References %%%

\begin{References}
    \printbibliography[heading=none] %Prints bibliography
\end{References}

\end{document}